\documentclass{elsart5p}

\usepackage{graphics}
\usepackage{graphicx}
\usepackage{amssymb}
\usepackage{amsmath}

\begin{document}

\begin{frontmatter}



\title{Effects of a Magnetic Field on Superconductivity and Quantum Criticality in Quasi-Two-Dimensional Systems with Dirac Electrons}
\thanks[CNPq]{This work has been
supported in part by CNPq and FAPERJ. ECM has been
partially supported by CNPq.}


\author[AA]{E. C. Marino
%\corauthref{Marino}
} and
%\ead{marino@if.ufrj.br}
\author[AA]{Lizardo H. C. M. Nunes}

\address[AA]{Instituto de F\'{\i}sica,  Universidade Federal do Rio de Janeiro,\\
Cx. P. 68528,  Rio de Janeiro-RJ 21941-972, Brazil}
%\address[BB]{Address 2}

%\corauth[Marino]{Corresponding author. Tel: +55 21  2562-7486 fax: +55 21 fax?}

\begin{abstract}
We study the effects of an external magnetic field on the
superconducting phase diagram
 of a quasi-two-dimensional system of
Dirac electrons at an arbitrary temperature. 
At zero temperature,
there is a quantum phase transition connecting a normal and a superconducting
phase, occurring at a critical line that corresponds to a magnetic
field dependent critical coupling parameter, 
which should be observed
in planar materials containing Dirac electrons, such as
$Cu_xTiSe_2$.
Moreover, 
the superconducting gap is obtained as a function of
temperature, magnetic field and coupling parameter ($\lambda_{\rm
R}$). From this, we extract the 
critical magnetic field $ B_{ c } $ as a function of the temperature.
For small values of $ B_{ c } $,
we obtain
a linear decay of the critical field, which
is similar to the behavior observed experimentally in the copper
doped dichalcogenide $Cu_xTiSe_2$ and also in intercalated graphite.
\end{abstract}


\end{frontmatter}

% main text

\section{Introduction}\label{int}

%A lot of attention has been devoted recently to
%quasi-two-dimensional condensed matter systems presenting a band
%structure such that the dispersion relation of the active electrons
%corresponds to the one of a relativistic massless particle. The
%kinematics of such electrons is described by a Dirac instead of a
%Schr\"odinger term in the hamiltonian \cite{diracelec,diracelec1}.
Among the materials presenting Dirac electrons as their elementary
excitations, there are a few, which have been intensely focused
lately. These are high-Tc cuprates \cite{htc}, 
%\cite{htc,htc1,htc2,htc3} 
graphene \cite{graf},
%\cite{graf,graf1,graf2}, 
carbon nanotubes \cite{carbnano},
%\cite{carbnano}
and transition metal dichalcogenides \cite{dcalc}.
%\cite{dcalc,dcalc1}

%A key issue, both from the basic and
%applied physics points of view is the analysis of the effects of an
%external magnetic field on the superconducting properties of a
%system. 
In the present work, we study the effects of an applied
constant magnetic field, perpendicular to a quasi-two-dimensional
superconducting system containing Dirac electrons.

The Lagrangian describing a quasi-two-dimensional
superconducting electronic system, containing two Dirac points,
in the presence of the external magnetic field 
along the c-axis, $B\hat{z}$,
is given by
\begin{eqnarray}
\mathcal{ L }
 & = &
{\rm i} \ \overline\psi_{ \sigma a} \left[ \hbar \partial_{ 0 } +
v_{ F }\gamma^{i } \left( \hbar  \partial_{ i } + {\rm i}
\frac{e}{c} A_{ i } \right) \right] \psi_{ \sigma a} 
\nonumber \\
& & +
 g
\left(
\psi^\dag_{1\uparrow a} \ \psi^\dag_{2\downarrow a}
+ \psi^\dag_{2\uparrow a} \ \psi^\dag_{1\downarrow a}
\right)
\left(
\psi_{2\downarrow b} \ \psi_{1\uparrow b}
+ \psi_{1\downarrow b} \ \psi_{2\uparrow b}
\right )
\nonumber \\
& & 
- \psi^{ \dagger }_{ \sigma a} 
\left( \mu_{ B }  \vec{ B } \cdot \vec{ \sigma } \right) 
\psi^{ \dagger }_{ \sigma a}
\, ,
\label{EqModel}
\end{eqnarray}
where  the electron creation operator will be
$\psi^\dag_{i,\sigma,a}$, where $i=1,2$ denotes the Dirac point and
$\sigma=\uparrow,\downarrow$, the z-component of the spin,
$a=1,...,N$ is an extra label, identifying the plane to which the
electron belongs, $A_{ i }$ is the vector potential corresponding to $\vec B$,
$\vec \sigma$ are Pauli matrices and $\mu_B$ is the Bohr magneton.
The second and third terms, respectively, contain the coupling of
the magnetic field to orbital and spin degrees of freedom. As in
\cite{npb}, we assume there is an effective superconducting
interaction whose origin will not influence the results of this
work. $g$ is the superconducting coupling constant, which is
supposed to depend on some external control parameter. In order to
make the lagrangian smooth, we define $g\equiv \lambda/N$. We use
the same convention for the Dirac matrices as in \cite{npb}.

Introducing a Hubbard-Stratonovitch complex scalar field
$\sigma = - g\ \left (\psi_{2\downarrow a} \ \psi_{1\uparrow a} +
\psi_{1\downarrow a}\ \psi_{2\uparrow a} \right ) $, 
where $\sigma^\dagger$ is a Cooper pair creation operator,
the Lagrangian,  in terms of $ \sigma $ becomes, 
\begin{equation}
\mathcal{ L } \left[ \Psi , \sigma \right]= - \frac{1}{ g }\
\sigma^{ * }  \sigma + \Psi_a^\dag \mathcal{ A } \Psi_a
\label{Eqsigmapsi}
\end{equation}
where
\begin{equation}
\mathcal{ A } =
\begin{pmatrix}
\tilde{ \partial_{ 0 } }  & - \tilde{ \partial_{ - } } & 0 & \sigma \\
- \tilde{ \partial_{ + } } & \tilde{ \partial_{ 0 } } & \sigma & 0 \\
0 & \sigma^{ * } & \tilde{ \partial_{ 0 } } & \tilde{ \partial_{ + } } \\
\sigma^{ * } & 0 & \tilde{ \partial_{ - } } &  \tilde{ \partial_{ 0
} } ,
\end{pmatrix}
\label{EqA}
\end{equation}
with $  \tilde{ \partial_{ 0 } } \equiv {\rm i } \ \left( \hbar
\partial_{0 } + \mu_{ B } B \right) $,
$  \tilde{ \partial_{ \pm } } \equiv {\rm i } \ v_{ F } \left( \hbar
\partial_{ \pm } + {\rm i } (e/c) A_{ \pm } \right) $ and
$ \partial_{ \pm } = \partial_{ 2 } \pm {\rm i } \ \partial_{ 1 } $.
The fermions are in the form of a Nambu field $ \Psi^\dag_a =
(\psi^\dag_{1\uparrow a}\ \psi^\dag_{2\uparrow a}\
\psi^\dag_{1\downarrow a}\ \psi^\dag_{2\downarrow a} ) $.

Integrating on the fermion fields , 
we obtain the effective action per plane for $\sigma$, 
namely
\begin{eqnarray}
S_{\rm eff}\left( |\sigma| , B \right) 
& = &
\int d^3 x \left( -\frac{ N
 }{ \lambda }|\sigma |^{ 2 }\right ) 
\nonumber \\
& & 
- {\rm i}N  \ln {\rm Det}\left[ \frac{
\mathcal{A} \left[ \sigma, B \right] }{
\mathcal{A}  \left[ \sigma= 0, B = 0 \right]}\right ] 
\, .
\label{EqSeff}
\end{eqnarray}

At $T = 0$, the renormalized 
effective potential per plane is
\begin{equation}
V_{ { \rm eff},R } \left( |\sigma| , B \right) = \frac{ |\sigma|^{ 2
} }{ \lambda_{ R } } - \frac{ f( B, \sigma_{ 0 } ) }{ \lambda_{ c }
} |\sigma|^{2} + \frac{ 2 }{ 3 \alpha } \left( |\sigma|^{2} + B
\kappa \right)^{ \frac{ 3 }{ 2 } } 
\, ,
\label{EqVeffR}
\end{equation}
where $ \kappa $ and $ \alpha $ denote  
$ v_{ F }^{ 2 } \hbar ( e / c ) $
and 
$ 2 \pi v_{ F }^{ 2 } $
respectively, $ \lambda_{ R } $ is the renormalized coupling
and $ \lambda_{ c } = 2 \alpha / 3 \sigma_{ 0 } $.

Studying the minima of the previous expression, 
the superconducting gap  $ \Delta \equiv |\langle 0|\sigma |0 \rangle| =  0 $ 
exists only for
\begin{equation}
\lambda_{R} <  \lambda_{ c } (B) = \lambda_{ c } \frac{ \sqrt{1+
\tilde{ B }}  }{1+ \frac{3}{2}  \tilde{ B }  - \frac{2}{3} \sqrt{
\tilde{ B }(1+ \tilde{ B }) } } 
\, ,
\label{EqSigma0Condition}
\end{equation}
where $ \tilde{ B } = B ( \kappa / \sigma_{ 0 }^{ 2 } ) $
and
$ \sigma_{ 0 } $ is an arbitrary finite scale,
and a quantum phase transition connecting a
normal and a superconducting phase occurs at the magnetic field
dependent quantum critical point $ \lambda_{ c } (B)$, given by
the above expression.
%(\ref{EqSigma0Condition}).

We turn now to finite temperature effects. Using a large $N$ expansion
and evaluating (\ref{EqSeff}) at $T\neq 0$, we find the effective potential,
whose minima provide a general expression for the
superconducting gap $\Delta(T,B)$, 
as a function of the temperature
and of the magnetic field,
\begin{eqnarray}
 \Delta^2(T,B) 
 & = &
\left \{
k_{ B } T  \cosh^{- 1 }
\left[ \frac{
e^{ \frac{ \sqrt{ \Delta_{ 0 }^{2 } + B \kappa } }{ k_{ B } T }
}  }{ 2 }
- \cosh\left( \frac{ \mu_{ B } B }{ k_{ B } T } \right)
\right]
\right \}^2
\nonumber \\
& &
- B \kappa
\, , 
\label{EqGapEquationSolution2}
\end{eqnarray}
where $ \Delta_{ 0 } \equiv \Delta( T = 0 ) $.

\begin{figure}[!ht]
\centerline {
\includegraphics[clip,width=0.30\textwidth 
,angle=-90 
]{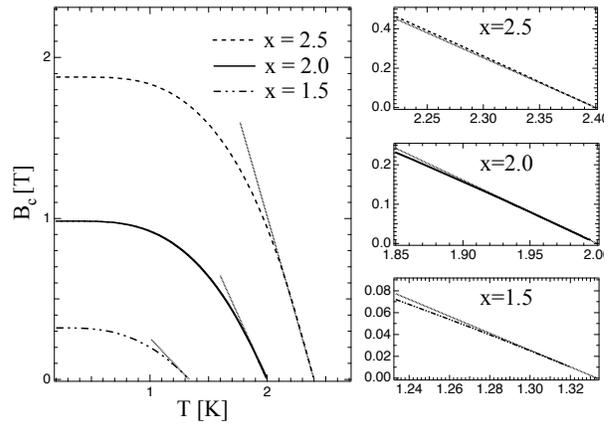} }
\caption{$ B $ vs. $ T $ phase diagram. 
The critical magnetic field $ B_{ c } $ as
a function of the temperature for several values of the
dimensionless coupling parameter $x \equiv \lambda_{\rm R} /
\lambda_{c} $. The dotted lines in the figure indicate the linear
behavior of $ B $ given by (\ref{EqBcrit}) as $ B \rightarrow 0 $.}
\label{FigBcxT}
\end{figure}

From (\ref{EqGapEquationSolution2}),
and using the fact that $ \Delta(T,B_{ c }) =  0 $
when the critical magnetic field $ B_{ c } $
is applied to the system, 
we can obtain  $ B_{c } $
vs. $ T$ phase diagram for the quasi-two-dimensional
superconducting Dirac electronic system. This is represented in Fig.
\ref{FigBcxT}. Particularly interesting is the linear behavior of
the critical magnetic field for $B \gtrsim 0$,
which is explicitly derived from (\ref{EqGapEquationSolution2})
in the small $B$ region:
\begin{equation}
B_c(T) \sim \frac{ 8 \ln 2 k_{ B }^{ 2 } }{ A \, \kappa } \, { T_{ c
}^{ 2 } ( 0 ) } \left( 1 - \frac{ T }{ T_{ c }( 0 ) } \right) 
\, ,
\label{EqBcrit}
\end{equation}
where
%\begin{equation}
$
A = 1 - ( 3 / 4 \ln 2 )
[ 
1 -( \lambda_{ c } / \lambda_{ R } ) 
]
$
%\, , \label{EqA}
%\end{equation}
and
%\begin{equation}
$
k_{ B } T_{ c } ( 0 ) =  
( 3 \sigma_{ 0 } / 4 \ln 2 )
[
1 - ( \lambda_{ c } / \lambda_{ R } )
]
$.
%\, . \label{EqA}
%\end{equation}

In Fig.  \ref{FigBcxT}, the linear behavior of the critical field
is indicated by the dotted lines for different values of the
dimensionless parameter $ x \equiv \lambda_{ R } / \lambda_{ c } $
(this is actually valid for $\lambda_{ R } <
13 \lambda_{ c } $, when $A$ is positive ). This differs from the
quadratic behavior predicted by BCS theory.

A linear decay of the critical field with the temperature similar to
the one obtained here has been experimentally observed in
intercalated graphite compounds \cite{intgraf} and also in the
copper-doped dichalcogenide $Cu_x Ti Se_2$  \cite{tmd}. Since
graphene and also the transition metal dichalcogenides are
well-known to possess Dirac electrons in their spectrum of
excitations, one is naturally led to wonder whether the presence of
such electrons could explain such a behavior of the critical field.

%%%%%%%%%%%%%%%%%%%%%%%
%%%%%%%%%%%%%%%%%%%%%%%%

%\section{Summary}\label{summary}

%\section{Acknowledgement}

%We would like to thank A.H.Castro Neto and B. Uchoa, for very
%helpful comments and conversations. This work has been supported in
%part by CNPq and FAPERJ. ECM has been partially supported by CNPq.

%%%%%%%%%%%%%%%%%%%%%%%%%%%%%%%%%%%%

%%%%%%%%%%%%%%%%%%%%%%%%%%%%%%%%%

\end{document}